\documentclass[prl,showpacs,floatfix,nofootinbib,superscriptaddress,twocolumn]{revtex4}

\usepackage{amsmath,amssymb} %
\usepackage{graphicx,times,xspace} %

\IfFileExists{srcltx.sty}{\usepackage[active]{srcltx}}

\newcommand{\adsurl}[1]{\href{#1}{ADS}}
\newcommand{\eprint}[1]{\href{http://arxiv.org/abs/#1}{#1}}

\newcommand{\pc}{\:\mathrm{pc}} 
\newcommand{\kpc}{\:\mathrm{kpc}} 
\newcommand{\mpc}{\:\mathrm{Mpc}} 
\newcommand{\parfrac}[2]{\left(\frac{#1}{#2}\right)} %
\renewcommand{\S}{\ensuremath{\mathcal{S}}\xspace} %
\newcommand{\totalsel}{289\xspace}
\begin{document}

\title{Universal properties of Dark Matter halos} 

\author{A.~Boyarsky}%
\affiliation{Ecole Polytechnique F\'ed\'erale de Lausanne, FSB/ITP/LPPC, BSP
  720, CH-1015, Lausanne, Switzerland} \affiliation{Bogolyubov Institute of
  Theoretical Physics, Kyiv, Ukraine} \author{A.~Neronov}%
\affiliation{INTEGRAL Science Data Center, Versoix
  and 
  Geneva Observatory, 
  Sauverny, Switzerland} \author{O.~Ruchayskiy}%
\affiliation{Ecole Polytechnique F\'ed\'erale de Lausanne, FSB/ITP/LPPC, BSP
  720, CH-1015, Lausanne, Switzerland} \author{I.~Tkachev}%
\affiliation{Institute for Nuclear Researches, Moscow, Russia}%
\date{November 17, 2009}

\begin{abstract}
  We discuss the universal relation between density and size of observed Dark
  Matter halos that was recently shown to hold on a wide range of scales, from
  dwarf galaxies to galaxy clusters. Predictions of $\Lambda$CDM N-body
  simulations are consistent with this relation. We demonstrate that this
  property of $\Lambda$CDM can be understood analytically in the secondary
  infall model. Qualitative understanding given by this model provides a new
  way to predict which deviations from $\Lambda$CDM or large-scale
  modifications of gravity can affect universal behavior and, therefore, to
  constrain them observationally.
\end{abstract}


\maketitle

{\bf Introduction}. The nature of Dark Matter (DM) is one of the most
fundamental puzzles in modern physics (see e.g.  \cite{Bergstrom:09} for a
recent review).  Resolution of this puzzle may potentially shed a light on
many fundamental physical issues, including particle physics beyond the
Standard Model, large-scale properties of gravity, etc.  The simplest Cold
Dark Matter (CDM) model is rather successful in explaining cosmological and
astronomical observational data at large distances~\cite{WMAP5cosmoParams}.
More complicated models, discussed in the literature, include ``warm'' DM
models \cite{Bode:00}, or models with modifications of large-scale properties
of gravity~\cite{DGP} or even of Newtonian dynamics \cite{Milgrom:83}.  Search
for deviations from the CDM predictions provides an important tool to
constrain the properties of unknown DM particles and to test fundamental laws
of gravity.

Comparison of predictions of DM models with observational data is complicated
by the fact that information about the properties of DM halos is usually
derived from the study of dynamics of ``luminous tracers'': stars and gas
particles (the only exception being gravitational lensing method). Physics of
the luminous baryonic matter in galaxies and galaxy clusters is very
nontrivial to be fully accounted for in theoretical and numerical modeling.
Thus, it is crucial to find the properties of DM distributions that do not
depend on the physics of the baryonic matter and, therefore, could be directly
compared with predictions of pure DM models.  Below, we argue that recently
discovered relation between characteristic density and size of the DM halos of
galaxies and galaxy clusters may provide an example of such a ``universal''
relation, and suggest its analytical understanding within the CDM model.

{\bf Universal properties of DM halos}. It was discussed for a long time
(see~\cite{Kormendy:04,Donato:09,Gentile:09a} and references therein) that
sizes and densities of observed galactic DM halos obey a universal scaling
relation. Namely, the DM central density $\bar \rho_{\rm C}$, averaged over a
region with the characteristic size $r_{\rm C}$ of the halo, scales roughly as
$r_{\rm C}^{-1}$.  This scaling law might appear surprising, if one takes into
account quite different histories of formation of various galaxies and their
different types. No clear qualitative understanding of the origin of
``dilution'' of the density of DM halos with the increase of the halo size was
proposed so far.

A different scaling relation between parameters of DM distribution was
suggested in~\cite{Boyarsky:09b} for still larger range of distance and mass
scales. The authors of~\cite{Boyarsky:09b} collected published data on the DM
distributions in \totalsel\ objects ranging from dwarf spheroidal galaxies
(dSphs) to galaxy clusters. The observational data were fit by different
density profiles: Navarro-Frenk-White (NFW)~\cite{Navarro:96},
(pseudo)-isothermal (ISO), Burkert (BURK)~\cite{Burkert:95} and on average 3
density profiles were collected for each object.  It was found that the
average ``column density'' of DM halos
\begin{equation}
 \label{eq:Sbar}
 \S = \frac{2}{r_{\rm C}^2} \int_0^{r_{\rm C}} r' dr' 
 \int_{-\infty}^{\infty} dz\, \rho(\sqrt{r'^2+z^2}) 
\end{equation}
slowly grows with the increase of the mass of the halo as 
\begin{equation}
 \label{eq:Sobs}
 \S \simeq 61\left[\frac{M_{200}}{10^{10}M_\odot}\right]^{0.21}\frac{M_\odot}{\mbox{pc}^2} 
\end{equation}
where $M_{200}$ is the measure of the halo mass (it is the mass contained
within a sphere of the radius $R_{200}$ where the average DM density equals to
$200\rho_{\rm crit}$ with $\rho_{\rm crit}$ being the critical density of the
Universe today). The parameter $r_{\rm C}$ in~(\ref{eq:Sbar}) is taken to be
$r_s$ of the NFW profile that fits the observed DM distribution. If the same
data (e.g.  rotation curves in galaxies or temperature profiles of galaxy
clusters) were fitted using a different DM profile, the scale $r_{\rm C}$ may
be uniquely expressed in terms of the parameters of that profile. The
column density~(\ref{eq:Sbar}) is insensitive to the type of DM density
profile used to fit the same observational data (see~\cite{Boyarsky:09b} for
details).

\begin{figure} %
  \centering
  \includegraphics[width=.8\linewidth]{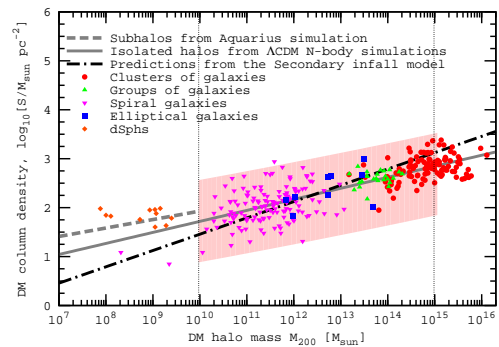} %
  \caption{Column density $\mathcal{S}$ as a function of halo mass $M_{200}$.
    The gray solid line is the relation~(\protect\ref{eq:Sobs}), coinciding
    with predictions of N-body simulations~\protect\cite{Maccio:08}, using the
    WMAP5 \protect\cite{WMAP5cosmoParams} cosmological parameters. The shaded
    region shows the $3\sigma$ scatter in the simulation data.  The vertical
    lines indicate the mass range probed by simulations.  The gray dashed line
    shows the results of the Aquarius simulation \protect\cite{Springel:08a}
    for sub-haloes. The black dashed-dotted line follows from the infall
    model.}
  \label{fig:S-M} 
\end{figure}

From Eq.~(\ref{eq:Sbar}) it follows that $\S \propto \bar\rho_{\rm C}r_{\rm
  C}$. Therefore, for cored profiles (ISO or BURK) the relation
(\ref{eq:Sobs}) is a generalization of the previously considered relation
$\bar\rho_{\rm C}r_{\rm C}\simeq const$~\cite{Kormendy:04,Donato:09}.
However, the quantity $\S$ is more universal, as it is defined for any (not
necessarily cored) DM profile. Therefore, in~\cite{Boyarsky:09b} $\S$ was
derived for \emph{any} type of DM density profile, used by observers to
describe the DM distribution in halos in the broad range of sizes
($0.2\kpc\lesssim r_{\rm C} \lesssim 2.5\mpc$) and masses
($10^8M_\odot<M_{200}<10^{16}M_\odot$).  The relation $\bar\rho_{\rm C}r_{\rm
  C}\simeq 141M_\odot \pc^{-2}$, derived in~\cite{Donato:09} for galaxies, and
$\S\sim 200 M_\odot \pc^{-2}$, proposed in the Ref.~\cite{Boyarsky:06c} for
galaxies and galaxy clusters, could serve as approximations of relation
(\ref{eq:Sobs}) at low- and high-mass ends of the data. This is clear from
Fig. \ref{fig:S-M} in which the data from~\cite{Boyarsky:09b} are presented
together with the fit (\ref{eq:Sobs}) (solid line).

The relation (\ref{eq:Sobs}) is derived from the analysis of data on DM halos
of structures with significantly different physics of the luminous matter.
(from stars in dwarf galaxies through gaseous disks at the outskirts of spiral
galaxies to hot intracluster medium gas in galaxy clusters).  The fact that
the properties of DM distributions in all these structures follow the same
scaling relation allows to conjecture that the observed scaling is a
``baryon-independent'' characteristics of DM halos.

A strong argument in favor of such a conjecture is given by agreement of the
relation (\ref{eq:Sobs}) derived from observational data with a similar
relation derived from \emph{pure DM} N-body simulations, performed in the
framework of the $\Lambda$CDM model~\cite{Maccio:08}.  The distributions of DM
particles in simulated halos are fitted by a density profile, from which $\S$
is computed using Eq.~(\ref{eq:Sbar}).  As discussed in \cite{Boyarsky:09b},
not only the trend $\S\sim M^{0.2}$, but also the scatter of individual
objects around the average (the pink shaded region in the Fig.~\ref{fig:S-M}
represents $3\sigma$ scatter of the data from the simulations of
Ref.~\cite{Maccio:08}) and a different slope for dwarf satellite galaxies
(gray dashed line) are reproduced extremely well.  The quantity $\S$ is
computed using the best-fit models obtained by different groups of observers
for each object and compared directly with the results of numerical
simulations, without questioning the suggested values of $M$ and $\S$.  The
agreement with N-body simulations confirms that, despite different nature and
quality of the observational data, the relation~\ref{eq:Sobs} is a physical
effect rather than a consequence of an observational bias.  Our goal here is
to explain this property of the $\Lambda$CDM model.

{\bf Secondary infall model}. In spite of apparent simplicity of relation
(\ref{eq:Sobs}) no qualitative explanation and/or analytical derivation of the
observed decrease of the average halo density with the increase of the halo
mass or size has been proposed so far. In what follows we show that this
effect could be readily understood within a semianalytical model of structure
formation, known as ``secondary infall'' model. Within this model, the column
density ${\cal S}$ is expected to change slowly with the increase of the
overall halo mass, $\S\sim M_{\rm halo}^{\kappa}$, with $\kappa\le 1/3$. Under
some simplifying assumptions (see below), the upper limit $\kappa=1/3$ is
achieved in structures which continue to grow at present.

The secondary infall
model~\cite{Gunn:72,Fillmore:84,Bertschinger:85a,Sikivie:95,Sikivie:96,Steigman:98}
(see also~\cite{DelPopolo:2009df} for a recent review) provides an analytic
description of the growth of the mass $M(t)$ and size $R(t)$ of DM halos in
the course of spherically symmetric accretion from cosmological matter flow.
One assumes that initial overdensity at the time $t_i$ is distributed in a
spherically symmetric way so that the initial mass distribution depends only
on the distance $r$ from the inhomogeneity center
\begin{equation}
  \label{eq:14}
  m(r,t_i)= \frac{4\pi}3\rho_M(t_i) r^3 + \delta m_i(r)
\end{equation}
where $\rho_M(t_i)=\left(6\pi Gt_i^2\right)^{-1}$ is the cosmological matter
density and $\delta m(r)$ is distance-dependent mass excess.\footnote{We
  neglect $\Lambda$-term for sufficiently early $t_i$.}

The entire DM
distribution is split into thin shells parameterized by their initial radii
$r_i=r(t_i)$.  Evolution of the radius $r(t)$ of each shell is governed by the Newtonian dynamics
\begin{equation}
  \label{eq:1}
  \frac{d^2 r}{dt^2} = - \frac{\partial U(r)}{\partial r}= -\frac{G m(t)}{r^2}\;,
\end{equation}
where $U(r)$ is the gravitational potential and $m(t)$ is the mass inside the
radius $r(t)$.  Initial velocities of the shells are assumed to follow Hubble
law $\dot r(t_i)=H(t_i)r_i$, with $H(t)$ being the expansion rate of the
Universe.

If $\delta m_i(r_i)>0$ expansion of the shell with given $r_i$ slows down
faster than it would do in the absence of central overdensity. A
straightforward calculation allows to find the time $t_*$ at which $\dot
r(t_*)=0$ and the shell reaches maximum (\emph{turnaround}) radius
$R_*=r(t_*)$
\begin{equation}
  \label{eq:28}
  t_* = \parfrac{\pi^2 R^3_*}{8 G m_i}^{1/2}\quad ;\quad R_* = r_i \frac{m_i}{\delta m_i (r_i)}\;.
\end{equation}
At any given moment of time $t$ one can define the boundary of the spherical
halo $R(t)$ as the turnaround radius of a shell for which $t=t_*$. From
Eq.(\ref{eq:28}) the mass within this radius, $M(t)=m_i$, is related to $R(t)$ via
\begin{equation}
  \label{eq:4}
  R(t)=\left(\frac{8GM(t)t^2}{\pi^2}\right)^{1/3}\;.
\end{equation}
Continuous growth of the size and mass of the spherical halos is similar to
spherically symmetric accretion onto a central mass concentration. In the
accretion theory, the ``sphere of influence'' of a body with the mass $M$ is
defined as the distance range within which the gravitational potential
$U=GM/R$ is larger than the specific kinetic energies of accreting particles.
In cosmological settings, the given kinetic energy of particles is that of the
Hubble flow, $K=(H(t)R)^2/2$. For each $R$ it decreases with time. This leads
to the continuous increase of the sphere of influence of the halo with the
mass $M$: $R\propto M^{1/3}/H^{2/3}$ and, respectively, to the growth of the
halo mass $M$, as Eq.~(\ref{eq:4}) demonstrates.  The accretion from the
cosmological flow could stop if the ``sphere of influence'' of a given DM halo
starts to intersect with the ``spheres of influence'' of the neighboring
halos.

At late stages of evolution of the Universe, the rate of accretion onto a DM
halo could diminish due to the influence of the cosmological constant onto the
dynamics of cosmological expansion in the direct vicinity of the halo.
Modification of the gravitational potential entering Eq. (\ref{eq:1})
\begin{equation}
  \label{eq:31}
 U(r) = -\frac{GM}r - \frac{4\pi G \rho_\Lambda}3 r^2
\end{equation}
by a cosmological constant (with the energy density $\rho_\Lambda$) leads to a
modification of the relation (\ref{eq:28}) between the $R_*$ and $t_*$ to
\begin{equation}
  \label{eq:5}
  t_* = \left(\frac{\pi^2 R_*^3}{8 GM}\right)^{1/2}
  f\left(\lambda\right)\quad;\quad \lambda = 4\pi \rho_\Lambda R_*^3/(3M)
\end{equation}
where $ f(\lambda) \equiv (2/\pi) \int^1_0 dx/{\sqrt{\left(x^{-1}
      -1\right)+\lambda(x^2 -1)}} $.  For the fixed $t_*$, the
relation~(\ref{eq:5}) has the form $\lambda^{1/2} f(\lambda) =
\mathrm{const}$.  This means that, for the fixed $t_*$,
$\lambda=\mathrm{const}$ and, therefore, the relation between $R_*$ and $M$,
implied by ~(\ref{eq:4}), still holds, although with $\approx 20\%$ different
normalization.

Once the shell passes the turnaround radius, it starts to oscillate around the
center of the halo. The shell will cross other shells and its dynamics will
become dependent on their movement. The amplitude of oscillations decreases
with time, as long as accretion from the cosmological matter flow continues.
The average (over the shells) radial density profile of the halo changes with
time.  In general, no analytic solution could be found for the dynamics of
each shell and the average halo density profile.  However, the study of the
halo growth simplifies if one considers initial perturbations with power-law
profiles
\begin{equation}
\label{self}
\frac{\delta M_i}{M_i}= \left(\frac{M_0}{M_i}\right)^{\epsilon}\;,
\end{equation}
where $\epsilon$ is the power-law index and $M_0$ is a normalization constant
(see~\cite[Sec III.D]{Sikivie:96} for the discussion of the relation of the
value of $\epsilon$ and initial spectrum of density perturbations).  In this
case the halo density profie evolves in a ``self-similar'' manner: its shape
can be expressed as
\begin{equation}
  \label{eq:45}
  \rho(r,t) = \frac{M(t)}{R^3(t)}\times F\left(\frac{r}{R(t)}\right)\;.
\end{equation}
The function $F(x)$ is the ``self-similar'' (time and mass scale independent)
DM radial density profile with a characteristic radius $x_{\rm C}$.
Self-similar solutions exist also in a more general infall models, considered
in~\cite{Sikivie:96}. One can introduce for each shell an angular momentum
$\ell$, while still keeping the problem spherically
symmetric~\cite{Sikivie:96}.  If the angular momentum $\ell = j {R_*^2}/{t_*}$
with $j=\mathrm{const}$, the initial conditions~(\ref{self}) still lead to a
self-similar solution.  The detailed shape of the function $F(x)$ can be found
numerically.  Its asymptotic behavior is given by~\cite{Sikivie:96}
\begin{equation}
  \label{eq:47}
  F(x) \propto \left\{
    \begin{aligned}
      x^{-\gamma}, && x \ll x_{\rm C}, && \gamma = \frac{9\epsilon}{3\epsilon+1}\\
      x^{-2}, && x\gg x_{\rm C}
    \end{aligned}\right.\;.
\end{equation}
The property of self-similarity allows to derive the behavior of the column
density $\S\propto \bar\rho_{\rm C}r_{\rm C} = \bar\rho_{\rm C}x_{\rm C} R(t)$
as a function of $M$:
\begin{equation}
  \label{eq:Sinfall}
  \S \propto \frac{M(t)}{R^2(t)} x_{\rm C} \propto M^{1/3}(t) t^{-4/3}
\end{equation}
where in the second relation we use the Eq.~(\ref{eq:4}) and $x_{\rm
  C}=\mathrm{const}$.
  
Equation (\ref{eq:Sinfall}) expresses the universal scaling relation between halo
column density and mass. It is a direct analog of relation (\ref{eq:Sobs})
within the secondary infall model. Notice, that it provides an explanation, at
least at the qualitative level, of the observed \emph{slow} growth of the DM
column density with the increase of the halo mass.  At the same time, the
Eq.~(\ref{eq:Sinfall}) contains, apart from the mass $M$, an additional
parameter $t$, the time during which the halo accretes from the cosmological
matter flow.

Notice that Eq.~(\ref{eq:4}) (or its analog~(\ref{eq:5})) implies that all
structures which accrete up to the time $t$ have the same average density
$\rho_R(t)=M(t)/\left[(4\pi/3)R^3(t)\right]$ equal to $\mathrm{const} \times
\rho_M(t)$ (the cosmological density at the time $t$).  For cosmologically
close objects (including those, analyzed in~\cite{Boyarsky:09b}) accretion
still continues, i.e. $t=t_0$ (the present age of the Universe).  If accretion
onto a halo has terminated at time $t_{\rm final}<t_0$, (this is the case, for
example, for the dwarf spheroidal satellites of the Milky Way), then according
to relation~(\ref{eq:Sinfall}) the DM column density of such objects will be
higher for the same mass.

For the density profile of the form~(\ref{eq:45}, \ref{eq:47}) the DM column
density~(\ref{eq:Sbar}) is given by
\begin{equation}
  \label{eq:42}
  \S =  \frac{\alpha}{x_{\rm C}}\parfrac{\pi^2}{8 G }^{2/3} 
  \frac{M^{1/3}(t_{\rm final})}{t_{\rm final}^{4/3}}\;,
\end{equation}
where $\alpha$ is a numerical coefficient which could be found from
numerically calculated profiles $F(x)$. Choosing a typical value for $x_{\rm
  C} \sim 0.02$~\cite{Sikivie:96}, one obtains $\S=133 M_\odot/$pc$^2$ for
$M\simeq 10^{12}M_\odot$. The relation~(\ref{eq:42}) (for $M=M_{200}$) with
this normalization is shown by the dashed-dotted line in Fig.~\ref{fig:S-M}.

{\bf Discussion.} The universal scaling relation between the mass and column
density of DM halos (\ref{eq:Sinfall}), derived within the simple
semianalytical model of self-similar secondary infall provides a reasonable
description of the data presented in Fig.~\ref{fig:S-M}. The scaling ${\cal
  S}\sim M^{1/3}$ is determined mostly by two properties of the model: {\it
  (i)} Average density inside the sphere of the radius $R(t)$ is the same for
all DM halos, independently of their mass.  It is equal, up to a constant, to
the cosmological matter density at the moment $t$, see Eq.  (\ref{eq:4}) and
(\ref{eq:5}). {\it (ii)} Radial density profiles of all halos are
approximately the same (the self-similarity condition~(\ref{eq:45})),
independently of their mass.

The statement {\it (i)} is just a general consequence of the physics of
accretion from cosmological matter flow.  The self-similarity property {\it
  (ii)} is a simplifying assumption. Although it is present in the simplest
solutions of the infall model (see~\cite{Sikivie:96}) and the data suggest
that the deviation from the self-similarity are not so big, it may or may not
hold in realistic structure formation. Validity of this assumption could be
verified via comparison with N-body simulations of structure formation. In the
simulations, density profiles of halos are usually fitted by two-parametric
NFW profiles, which may be characterized, for example, by the scales $r_{\rm
  C}=r_s$ and $R_{200}$.  The NFW profiles which arise in the simulations are
known (see e.g.~\cite{Maccio:08}) to satisfy the relation
\begin{equation}
  \label{eq:01}
c_{200} \propto M_{200}^{-0.1},
\end{equation}
where $ c_{200}=R_{200}/r_s \propto1/x_{\rm C}$. The dependence of $x_C$ upon
mass scale is rather weak and the density profiles predicted by N-body
simulations do form, to a good approximation, a self-similar subset of all NFW
profiles.
 
The small deviation from self-similarity, described by the
Equation~(\ref{eq:01}), explains a somewhat better fit to the data provided by
the N-body simulations.  Expressing $x_{\rm C}$ from Eq.~(\ref{eq:01}) and
substituting it into~(\ref{eq:42}) one finds ${\S}\sim M^{0.23}$, rather than
${\cal S}\sim M^{1/3}$, which is closer to the ${\S}(M)$ scaling relation
(\ref{eq:Sobs}) derived from the observational data.

In the frameworks of the infall models deviations from self-similarity arise
from slight dependence of $\epsilon$ in initial conditions~(\ref{self}) and/or
the angular momentum $\ell$ on the mass scale~\cite{Sikivie:96}.  Finally,
dependence of the ${\cal S}(M)$ scaling (\ref{eq:42}) on $t_{\rm final}$
introduces additional correction for structures in which $t_{\rm final}<t_0$.
More precise understanding of the deviations from the simplest approximation,
discussed above, together with better quality of the data, would allow to
observationally constrain the details of cosmological DM halo formation
process.  The detailed analysis of the small deviations from the self-similar
infall model is, however, beyond the scope of the present Letter.

{\bf Conclusions}. The universal $\S$--$M$ scaling in the DM halos, found in
the observational data and in pure DM N-body simulations can be analytically
understood in the secondary infall model. It seems to be insensitive to the
presence of baryons and to the details of DM density distributions. This shows
that this relation has pure DM origin (in contrast with its interpretation in
favor of Modified Newtonian Dynamics, as discussed e.g.
in~\cite{Gentile:09a}).

Qualitative understanding of the scaling between the density and size of DM
halos discussed in this Letter, opens a possibility to estimate expected
modification of the scaling over a broad parameter space of alternative models
of DM (such as e.g.~\cite{Bode:00}) and/or large-scale modifications of
gravity (see~\cite{Boyarsky:10a}) and, therefore, distinguish between
different classes of such models. In this respect, the simplified analytical
approach of the secondary infall model provides a valuable alternative to
resource consuming N-body simulations.

The predictions of the secondary infall model can be tested in the Local
Group~\cite{Steigman:98}, using e.g. the data on the local Hubble
flow~\cite{Karachentsev:08a}. Our results also motivate dedicated astronomical
search, especially for the isolated galaxies with the masses below $10^{10}
M_\odot$ (two pink downward triangles in Fig.~\ref{fig:S-M}) and systematic
studies of DM content in galaxy clusters, as well as a uniform analysis of all
observations collected in~\cite{Boyarsky:09b}.  This would allow to reduce
systematic errors in the data, determine more precisely the slope in
relation~(\ref{eq:Sobs}) and find possible deviations from the pure CDM model.

This work was supported in part by the SCOPES project No.  IZ73Z0\_128040 and
by the Swiss National Science Foundation.

\let\jnlstyle=\rm\def\jref#1{{\jnlstyle#1}}\def\aj{\jref{AJ}}
  \def\araa{\jref{ARA\&A}} \def\apj{\jref{ApJ}\ } \def\apjl{\jref{ApJ}\ }
  \def\apjs{\jref{ApJS}} \def\ao{\jref{Appl.~Opt.}} \def\apss{\jref{Ap\&SS}}
  \def\aap{\jref{A\&A}} \def\aapr{\jref{A\&A~Rev.}} \def\aaps{\jref{A\&AS}}
  \def\azh{\jref{AZh}} \def\baas{\jref{BAAS}} \def\jrasc{\jref{JRASC}}
  \def\memras{\jref{MmRAS}} \def\mnras{\jref{MNRAS}\ }
  \def\pra{\jref{Phys.~Rev.~A}\ } \def\prb{\jref{Phys.~Rev.~B}\ }
  \def\prc{\jref{Phys.~Rev.~C}\ } \def\prd{\jref{Phys.~Rev.~D}\ }
  \def\pre{\jref{Phys.~Rev.~E}} \def\prl{\jref{Phys.~Rev.~Lett.}}
  \def\pasp{\jref{PASP}} \def\pasj{\jref{PASJ}} \def\qjras{\jref{QJRAS}}
  \def\skytel{\jref{S\&T}} \def\solphys{\jref{Sol.~Phys.}}
  \def\sovast{\jref{Soviet~Ast.}} \def\ssr{\jref{Space~Sci.~Rev.}}
  \def\zap{\jref{ZAp}} \def\nat{\jref{Nature}\ } \def\iaucirc{\jref{IAU~Circ.}}
  \def\aplett{\jref{Astrophys.~Lett.}}
  \def\apspr{\jref{Astrophys.~Space~Phys.~Res.}}
  \def\bain{\jref{Bull.~Astron.~Inst.~Netherlands}}
  \def\fcp{\jref{Fund.~Cosmic~Phys.}} \def\gca{\jref{Geochim.~Cosmochim.~Acta}}
  \def\grl{\jref{Geophys.~Res.~Lett.}} \def\jcp{\jref{J.~Chem.~Phys.}}
  \def\jgr{\jref{J.~Geophys.~Res.}}
  \def\jqsrt{\jref{J.~Quant.~Spec.~Radiat.~Transf.}}
  \def\memsai{\jref{Mem.~Soc.~Astron.~Italiana}}
  \def\nphysa{\jref{Nucl.~Phys.~A}} \def\physrep{\jref{Phys.~Rep.}}
  \def\physscr{\jref{Phys.~Scr}} \def\planss{\jref{Planet.~Space~Sci.}}
  \def\procspie{\jref{Proc.~SPIE}} \let\astap=\aap \let\apjlett=\apjl
  \let\apjsupp=\apjs \let\applopt=\ao

\end{document}